\documentclass[10pt]{article}
\usepackage{amsmath,amssymb,amsfonts}
\usepackage{bbm,bm}
\usepackage{fullpage}

\title{\Large\bf QUATERNIONIC SCALAR FIELD \\IN THE REAL HILBERT SPACE \vspace{2mm}}
\author{{\bf Sergio Giardino\footnote{\tt sergio.giardino@ufrgs.br}}\\
\small \it Departamento de Matem\'atica Pura e Aplicada \\
\small \it Universidade Federal do Rio Grande do Sul (UFRGS)\\
\small \it Caixa Postal 15080, 91501-970  Porto Alegre RS \\
\small \it Brazil}
\begin{document}
\date{} 
\maketitle

\begin{abstract}
\noindent Using the complex Klein-Gordon field as a model, we quantize the quaternionic scalar field in the real Hilbert space. The lagrangian formulation has accordingly been obtained, as well as the hamiltonian formulation, and the energy and charge operators. Conversely to the complex case, the quaternionic quantization admits two quantization schemes, concerning either two or four components. Therefore, the quaternionic field permits a richer structure of states, if compared to the complex scalar field case. Moreover, the quaternionic theory admits as a further novel feature a non-associative algebraic structure in their complex components, something not observed in the complex case.
%
%
%
%
\end{abstract}

\maketitle

%
%
%
%

\maketitle
\section{\;\bf Introduction\label{I}}

On May 16, 1960, Theodore  Maiman achieved an experimental device able to generate a burst of coherent light:
the laser beam  \cite{Maiman}. Because of the many possible applications, Maiman once said in a interview to {\it The New York Times} that ''a laser is a solution seeking a problem''. In this article, we entertain a mathematical solution that seeks physical problems to solve: the quaternionic number ($\mathbbm H$).
Although much older than the laser (quaternions were discovered by William Rowan Hamilton in 1843), the application of quaternions in Physics is still not very popular, partially because it is not established what kind of physical problem quaternions may solve, but principally because we cannot clearly identify what problems cannot be solved without them. In other words,  quaternionic numbers will be  physically relevant only whether the resort to them be crucial to clarify a particular physical problem.

Further information about quaternions will be given in the next section, but we anticipate  that quaternions are hyper-complex numbers comprising four real components that can be used to model several applications of physical interest, such as Lorentz transformations \cite{Teli:1980mr,Morita:1985rx,DeLeo:2000ik,Morita:2007vc}, vector calculus \cite{Gough:1984qsh}, the Dirac equation \cite{Edmonds:1977fp,Berezin:1981tq,Arbab:2017ymu}, and several other \cite{weiss:1941aqr,Lambek:1995ihp,Frenkel:2007cm,Venancio:2020qqd}. However, these are simply different ways of solving well-known problems. An authentic example of a quaternionic physical proposal that cannot be understood as to rephrase the  complex physical theories concerns the formulation of quantum mechanics using quaternions. There are mainly two proposals to built a quaternionic quantum mechanics ($\mathbbm H$QM): the anti-hermitian theory \cite{Adler:1995qqm} is built over a quaternionic Hilbert space, and is the most studied proposal, despite the serious shortcoming of breaking down the Ehrenfest theorem ({\it cf.} Section 4.4 of \cite{Adler:1995qqm}). In recent past, a variant of  $\mathbbm H$QM has been developed over the real Hilbert space \cite{Giardino:2018rhs,Giardino:2018lem}, and the hope for a coherent quaternionic quantum theory has been renewed. The real Hilbert space approach to $\mathbbm H$QM restores Ehrenfest's theorem and the classical limit, and simple quaternionic quantum solutions that were never obtained in the anti-hermitian theory have been achieved \cite{Giardino:2016xap,Giardino:2017yke,Giardino:2017pqq,Giardino:2019xwm,Giardino:2020ztf,Giardino:2020cee,Giardino:2021ofo}. 
The $\mathbbm H$QM has also been extended to the relativistic theory, and quaternionic versions of the Klein-Gordon \cite{Giardino:2021lov}, and of the Dirac equations \cite{Giardino:2021mjj} have been successfully obtained. In the present article, we further extend this theory to the quantum scalar field using the quantization of the complex scalar field as a model, hence obtaining the quaternionic field, their lagrangian expression, and the quaternionic charge operator. 

Although the quantization of the quaternionic scalar field presented in this article is original, there is a collection of research works where quaternions are used as an alternative way to describe quantum field theories  in a complex Hilbert space \cite{Pushpa:2010sw,Afonso:2012zj,DAdda:2016rgl}. Moreover, there are also authentic  applications  where  the quaternionic Hilbert space has been used \cite{Horwitz:1984ep,Adler:1985uh,Nash:1985xf,Nash:1987sj,DeLeo:1991mi,DeLeo:1996ger,DeLeo:1996ac,Giardino:2012ti,Giardino:2015ola,Giardino:2015dza,Steinberg:2020xvf}, and the research reported in this article also belongs to this investigation effort, however using the real Hilbert space approach.

There are several features of the quaternionic theory. First of all, the complex components are independently quantized, and have also independent energies, something that is not observed in the complex scalar field. Another interesting characteristic is a non-associative character that has already been previously observed in quaternionic quantum mechanics. 
 The quantization method is simple, and takes advantage of the complex components of the field, thus enabling us to obtain a the lagrangian, and the quantized theory without resort to the definition of a quaternionic derivative. 
\section{\;\bf The Klein-Gordon quaternionic equation\label{F}}

We do not provide a survey on quaternions in this article, and introductory material can be found elsewhere
\cite{Morais:2014rqc,Rocha:2013qtt,Garling:2011zz,Ward:1997qcn}. However, in order to establish the notation, we recall that a quaternion $\,q\,$ is a four-dimensional hyper-complex number that can be written as
\begin{equation}\label{f01}
 q=x_0 + x_1 i + x_2 j + x_3 k, \qquad\mbox{where}\qquad x_0,\,x_1,\,x_2,\,x_3\in\mathbbm{R},\qquad i^2=j^2=k^2=-1.
\end{equation}	
The imaginary units $\,i,\,j\,$ and $\,k\,$ are anti-commutative, and satisfy the general law
\begin{equation}\label{f02}
e_a e_b =\epsilon_{abc}e_c-\delta_{ab},
\end{equation}
where $e_a$ represents the imaginary units, $\,a,\,b,\,c=\{1,\,2,\,3\},\,$ the symbol $\,\epsilon_{abc}\,$ is the Levi-Civit\'a tensor, and $\,\delta_{ab}\,$ stands for the Kronecker delta. The extended four-dimensional real notation for quaternions (\ref{f01}) can be turned into the symplectic two-dimensional  complex notation, such as
\begin{equation}\label{f03}
q=z_0+z_1j,\qquad\mbox{where}\qquad z_0=x_0+x_1i\qquad\textrm{and}\qquad z_1=x_2+x_3i.
\end{equation}
Using this notation, we introduce the quaternionic wave function
\begin{equation}\label{f04}
\Phi=\cos\Theta\,\phi^{(0)}+\sin\Theta \,\phi^{(1)}\,j,
\end{equation}
where $\,\phi^{(\alpha)}\,$ are complex functions, $\,\alpha=\{0,\,1\},\,$ and $\Theta$ is a real function. Using the wave function (\ref{f04}), and the Klein-Gordon equation
\begin{equation}\label{f05}
\Big(\Box+m^2\Big)\Phi=0,\qquad\quad\mbox{where}\qquad\qquad\Box=\partial_\mu\partial^\mu
\end{equation} 
is the usual D'Alembertian operator, we obtain the quaternionic equation
\begin{eqnarray}
&&\nonumber\cos\Theta\Big(\Box+m^2-\partial_\mu\Theta\,\partial^\mu\Theta\Big)\phi^{(0)}-\sin\Theta\Big(\Box\Theta+2\,\partial_\mu\Theta\,\partial^\mu\Big)\phi^{(0)}+\\
&&+\Big[\sin\Theta\Big(\Box+m^2-\partial_\mu\Theta\,\partial^\mu\Theta\Big)\phi^{(1)}+\cos\Theta\Big(\Box\Theta+2\,\partial_\mu\Theta\,\partial^\mu\Big)\phi^{(1)}\Big]j=0,
\end{eqnarray}
that can be rewritten in terms of their complex components as
\begin{eqnarray}\label{f06}
\Big(\Box	+m^2-\partial_\mu\Theta\,\partial^\mu\Theta\Big)\phi^{(\alpha)}=0\,&&\\
\label{f07}
\,\partial_\mu\left[\,\partial^\mu\Theta\, \left(\phi^{(\alpha)}\right)^2\,\right]=0.&&
\end{eqnarray}
The solutions for these equations have already been studied \cite{Giardino:2021lov}, where several consistency tests and interesting features were considered. The equations of motion (\ref{f06}) can be simplified after the transformations
\begin{equation}
\phi^{(\alpha)}\to e^{i\Theta}\phi^{(\alpha)},
\end{equation}
and considering (\ref{f07}) as constraints. The simplest solution is such as
\begin{equation}\label{f08}
\phi^{(\alpha)}=\exp\big[\pm i k^{(\alpha)}_\mu x^\mu\big],\qquad\mbox{and}\qquad\Theta=\theta_\mu x^\mu +\Theta_0,
\end{equation}
where $\,k^{(\alpha)\mu}\,$ and $\,\theta^\mu\,$ are real four vectors and $\,\Theta_0\,$ is a real constant. These solutions and equations (\ref{f06}) immediately give 
\begin{equation}\label{f09}
k^{(\alpha)}_\mu k^{(\alpha)\mu}+\theta_\mu\theta^\mu=m^2,
\end{equation}
what alternatively means that
\begin{equation}
k^{(0)}_\mu k^{(0)\mu}=k^{(1)}_\mu k^{(1)\mu}.
\end{equation}
The constraints (\ref{f07}) additionally give
\begin{equation}\label{f10}
\theta_\mu k^{(\alpha)\mu}=0.
\end{equation}
Interpreting the constraints (\ref{f09}-\ref{f10}) in terms of the the effective four momenta $p^{(\alpha)\mu}$, 
\begin{equation}\label{f11}
p^{(\alpha)}_\mu p^{(\alpha)\mu} =m^2,\qquad\mbox{where}\qquad p^\mu=k^\mu \pm\theta^\mu,
\end{equation}
the suitability of the quaternionic field for special relativity is ascertained in terms of two energy relations.
As we decoupled the quaternionic problem in terms complex equations, we can emulate the quantum field theory in the complex Hilbert space to get the second quantization of the quaternionic field, something that will be done in the next section.

\section{The four components quantization}
It is necessary to highlight that the quantization of the quaternionic scalar field is not unique, and the procedure may follow either a two components or a four components quantization procedure. Let us first study the four components procedure defining four complex fields such as
\begin{equation}\label{fc01}
\Phi^{(1)}=e^{i\Theta}\phi^{(0)},\qquad 
\Phi^{(2)}=e^{-i\Theta}\phi^{(0)},\qquad
\Phi^{(3)}=e^{i\Theta}\phi^{(1)},\qquad
\Phi^{(4)}=e^{-i\Theta}\phi^{(1)},
\end{equation}
where the quaternionic field will be
\begin{equation}\label{fc02}
\Phi=\frac{1}{2}\left(\Phi^{(1)}+\Phi^{(2)}\right)+\frac{1}{2i}\left(\Phi^{(3)}-\Phi^{(4)}\right)j.
\end{equation}
We notice that the constraint
\begin{equation}
 \Phi^{(1)}\Phi^{(4)}=\Phi^{(2)}\Phi^{(3)}
\end{equation}
is already considered within the momentum constraint (\ref{f11}).
Accordingly, we write the the lagrangian density in terms of a sum of lagrangian densities of the complex components,
\begin{equation}\label{fc03}
\mathcal L=\frac{1}{4}\sum_{a=1}^4\mathcal L^{(a)},
\qquad\qquad\mbox{where}\qquad\qquad
\mathcal L^{(a)}=\partial_\mu\Phi^{(a)}\,\partial^\mu\Phi^{(a)\dagger}-m^2\Phi^{(a)}\Phi^{(a)\dagger},
\end{equation}
as well as the hamiltonian density
\begin{equation}\label{fc04}
\mathcal H=\frac{1}{4}\sum_{a=1}^4\mathcal H^{(a)},
\qquad\qquad\mbox{where}\qquad\qquad
\mathcal H^{(a)}=\Pi^{(a)\dagger}\Pi^{(a)}+\bm\nabla\Phi^{(a)\dagger}\bm{\cdot\nabla}\Phi^{(a)}+m^2\Phi^{(a)\dagger}\Phi^{(a)}.
\end{equation}
The factor $1/4$ in (\ref{fc03}-\ref{fc04}) is necessary in order to compensate the factor that could be included in the definition (\ref{fc01}), but that were kept in (\ref{fc02}). Correspondingly, the momentum operator is obtained from the usual complex definition
\begin{equation}\label{fc05}
\Pi^{(a)}=\frac{\partial\mathcal L}{\partial\dot\Phi^{(a)\dagger}}.
\end{equation}
The formulation of the quaternionic problem in terms of their complex components  is the essential idea behind the method for obtaining the quaternionic scalar field theory used in this article, and the previous attempts to obtain a $\mathbbm H$QFT \cite{Adler:1995qqm,Adler:1985uh,DeLeo:1991mi} were unsuccessful because  this simple proposition was not introduced.
Therefore, the quantized quaternionic quantum field is obtained from the straightforward quantizaton of their complex components, so that
\begin{equation}\label{fc06}
\widehat\Phi^{(a)}\big( x\big)=\int\frac{d^3p^{(a)}}{\sqrt{(2\pi)^3\big|2p^{(a)}_0\big|}}\Big\{\exp\big[-ip^{(a)}_\mu x^\mu\big]\widehat a^{(a)}+ \exp\big[ip^{(a)}_\mu x^\mu\big] \widehat b^{(a)\dagger}\Big\},
\end{equation}
where
\begin{equation}\label{fc07}
p^{(1)\mu}=k^{(1)\mu}+\theta^\mu,\qquad
p^{(2)\mu}=k^{(2)\mu}-\theta^\mu,\qquad
p^{(3)\mu}=k^{(3)\mu}+\theta^\mu,\qquad
p^{(4)\mu}=k^{(4)\mu}-\theta^\mu.
\end{equation}
The annihilation operators,
\begin{equation}\label{fc08}
\widehat a^{(a)}=\widehat a\big(\bm p^{(a)}\big)\qquad\mbox{and}\qquad\widehat b^{(a)}=\widehat b\big(\bm p^{(a)}\big),
\end{equation}
as  well as the creation operators $ \widehat a^{(a)\dagger}$ and $ \widehat b^{(a)\dagger}$,  satisfy the equal-time  commutation relation
\begin{equation}\label{fc09}
\left[\widehat a\big(\bm p^{(a)}\big)	,\,\widehat a^\dagger\big(\bm p'^{(b)}\big)\right]_{x_0=y_0}=\left[\widehat b\big(\bm p^{(a)}\big)	,\,\widehat b^\dagger\big(\bm p'^{(b)}\big)\right]_{x_0=y_0}=\delta^{ab}\delta\left(p^{(a)}-p'^{(b)}\right),
\end{equation} 
while the remaining commutation relations are zero. The creation and annihilation operators ascribe the four momentum $\,p^{(a)\mu}\,$ to each quantum state, and hence the  contributions of $\Theta$ and $\phi^{(\alpha)}$ to the quantum states are all considered, so that
\begin{equation}\label{fc10}
a^{(1)\dagger}\big|0\big\rangle=\big|\theta,\,k^{(0)}\big\rangle,\qquad
a^{(2)\dagger}\big|0\big\rangle=\big|-\theta,\,k^{(0)}\big\rangle,\qquad
a^{(3)\dagger}\big|0\big\rangle=\big|\theta,\,k^{(1)}\big\rangle,\qquad
a^{(4)\dagger}\big|0\big\rangle=\big|-\theta,\,k^{(1)}\big\rangle.
\end{equation}
 Accordingly, the equal-time commutation relations of the fields are the usual complex, so that
\begin{eqnarray}
\nonumber\Big[\widehat\Phi^{(a)}(x),\,\widehat\Phi^{(b)}(y)\Big]_{x_0=y_0}&=&\Big[\widehat\Phi^{(a)}(x),\,\widehat\Phi^{(b)\dagger}(y)\Big]_{x_0=y_0}=\Big[\widehat\Phi^{(a)\dagger}(x),\,\widehat\Phi^{(b)\dagger}(y)\Big]_{x_0=y_0}=0\\
\label{fc11}\Big[\widehat\Pi^{(a)}(x),\,\widehat\Pi^{(b)}(y)\Big]_{x_0=y_0}&=&\Big[\widehat\Pi^{(a)}(x),\,\widehat\Pi^{(b)\dagger}(y)\Big]_{x_0=y_0}=\Big[\widehat\Pi^{(a)\dagger}(x),\,\widehat\Pi^{(b)\dagger}(y)\Big]_{x_0=y_0}=0\\
\nonumber\Big[\widehat\Phi^{(a)}(x),\,\widehat\Pi^{(b)\dagger}(y)\Big]_{x_0=y_0}&=&\Big[\widehat\Phi^{(a)\dagger}(x),\,\widehat\Pi^{(b)}(y)\Big]_{x_0=y_0}=i\,\delta^{ab}\,\delta^3(x-y).
\end{eqnarray}
Using the above results, we can connect the previous results to the wave solution of the Klein-Gordon equation, so that
\begin{equation}\label{fc12}
\big\langle p^{(\alpha)}\big|\widehat\Phi^{(\beta)}\big\rangle=\delta^{\alpha\beta}\frac{1}{\sqrt{(2\pi)^3 \big|2p^{(\alpha)}_0\big|}}\exp\left[-ip^{(\alpha)}_\mu x^\mu\right]
\end{equation}
Defining the quaternionic state operators
\begin{equation}\label{fc13}
\widehat a=\sum_{\alpha=1}^4\sqrt{\big|2p^{(\alpha)}_0\big|}\widehat a^{(\alpha)},\qquad\mbox{and}\qquad
\widehat b=\sum_{\alpha=1}^4\sqrt{\big|2p^{(\alpha)}_0\big|}\widehat b^{(\alpha)},
\end{equation}
we obtain the quaternionic Klein-Gordon wave function (\ref{f04})
\begin{equation}\label{fc14}
\big\langle p\big|\widehat\Phi\big\rangle=\frac{1}{\sqrt{(2\pi)^3 }}\Phi
\end{equation}
where 
\begin{equation}\label{fc15}
\langle p |=\langle 0|\widehat a,
\end{equation}
and
\begin{equation}\label{fc16}
\big|\widehat\Phi\big\rangle=\Big[\frac{1}{2}\left(\widehat\Phi^{(1)\dagger}+\widehat\Phi^{(2)\dagger}\right)+\frac{1}{2i}\left(\widehat\Phi^{(3)\dagger}-\widehat\Phi^{(4)\dagger}\right)j\,\Big]\Big|\,0	\Big\rangle.
\end{equation}
Therefore, we can recover the quaternionic wave function from the quantized complex components of the field, and this is a fundamental consistency test to the method. 
Finally, we obtain the energy operator
\begin{equation}\label{fc17}
:\widehat H:=\frac{1}{4}\sum_{a=1}^4\int d^3p^{(\alpha)}\big|p_0^{(\alpha)}\big|\left[\widehat a^{(\alpha)\dagger}\widehat a^{(\alpha)}+\widehat b^{(\alpha)\dagger}\widehat b^{(\alpha)}\right],
\end{equation}
and the charge operator
\begin{equation}\label{fc18}
\widehat Q=\frac{1}{4}\sum_{a=1}^4\int d^3p^{(\alpha)}\left[\widehat a^{(\alpha)\dagger}\widehat a^{(\alpha)}-\widehat b^{(\alpha)\dagger}\widehat b^{(\alpha)}\right],
\end{equation}
and the quantization of the field is thus complete. 
As a concluding remarks to this section, we firstly enumerate the four possible wave functions alternative to (\ref{fc16})
\begin{equation}\label{fc19}
\big|\widehat\Phi\big\rangle
\left\{
\begin{array}{l}
\Big[\frac{1}{2}\left(\widehat\Phi^{(1)\dagger}+\widehat\Phi^{(2)\dagger}\right)+\frac{1}{2i}\left(\widehat\Phi^{(3)\dagger}-\widehat\Phi^{(4)\dagger}\right)j\,\Big]\Big|\,0	\Big\rangle\\ \\
\Big[\frac{1}{2}\left(\widehat\Phi^{(1)}+\widehat\Phi^{(2)}\right)+\frac{1}{2i}\left(\widehat\Phi^{(3)\dagger}-\widehat\Phi^{(4)\dagger}\right)j\,\Big]\Big|\,0	\Big\rangle\\ \\
\Big[\frac{1}{2}\left(\widehat\Phi^{(1)\dagger}+\widehat\Phi^{(2)\dagger}\right)+\frac{1}{2i}\left(\widehat\Phi^{(3)}-\widehat\Phi^{(4)}\right)j\,\Big]\Big|\,0	\Big\rangle\\ \\
\Big[\frac{1}{2}\left(\widehat\Phi^{(1)}+\widehat\Phi^{(2)}\right)+\frac{1}{2i}\left(\widehat\Phi^{(3)}-\widehat\Phi^{(4)}\right)j\,\Big]\Big|\,0	\Big\rangle.
\end{array}
\right.
\end{equation}
In principle, there could be additional complex states, such as $\Phi^{(1)}+\Phi^{(2)\dagger}$, but these states cannot generate a wave function in the form (\ref{f04}) from (\ref{fc14}), and can be considered composite states that must be expanded in a base of the Hilbert space. Finally, the operators (\ref{fc19}) cannot be used to build a quaternionic commutation algebra in the same fashion as (\ref{fc11}) because the imaginary unit $\,j\,$ introduces an additional anti-commuting factor. On the other hand, the quaternionic imaginary unit introduces a non-associative feature of the quaternionic field that we explore in the next section.
\subsection{Non-associative feature}
The non-associative character of the quaternionic quantum field is already known in quaternionic quantum mechanics \cite{Giardino:2012ti}, and in this section we will identify it in the quaternionic field theory. Let us recall that the simplest characterization of a non-associative multiplicative algebra employs the associator,  defined as
\begin{equation}\label{na01}
\big(x,\,y,\,z\big)=x\big(yz\big)-\big(xy\big)z.
\end{equation}
Additionally, assuming the operator algebra (\ref{fc11}) as an alternative algebra \cite{Schafer:1966naa}, where the permutation of the elements in a associator is proportional to the Levi-Civit\'a symbol, and the Moufang identities hold. In this algebra, it holds the identity between the associator and the commutator
\begin{eqnarray}
\nonumber \left(\Phi^{(a)},\,\Pi^{(b)\dagger},\,j\right)&=&\left[\,\Phi^{(a)}\Pi^{(b)\dagger},\,j\,\right]\\
\nonumber &=&\left[\,\Pi^{(b)\dagger}\Phi^{(a)}+i\delta^{ab}\delta^3(x-y),\,j\,\right]\\
\nonumber &=&\left(\Pi^{(b)\dagger},\,\Phi^{(a)},\,j\right)+\delta^{ab}\delta^3(x-y)\big[\,i,\,j\,\big],
\end{eqnarray}
therefore,
\begin{equation}
\nonumber \left(\Phi^{(a)},\,\Pi^{(b)\dagger},\,j\right)= k\,\delta^{ab}\delta^3(x-y),
\end{equation}
where the alternative property of the alternator, the Moufang identities, and $ij=k$ for alternative algebras have been used. Interestingly, the the associator between annihilation and creation operators is zero because their commutation relation (\ref{fc09}) do not depend on the imaginary unit. The  implications of the non-associative feature in quantum field theory concerns an intriguing for future research.
\section{The two components quantization}
This quantization method is simpler than the previous one, because $\,\Theta=\Theta_0.\,$ Using the symplectic notation (\ref{f04}), the complex equations of motion (\ref{fc06}) can be obtained from the lagrangian density
\begin{equation}\label{tc00}
\mathcal L=\cos^2\Theta_0\, \mathcal L^{(0)}+\sin^2\Theta_0\,\mathcal L^{(1)},
\end{equation}
where
\begin{equation}
\mathcal L^{(\alpha)}=\partial_\mu\phi^{(\alpha)}\,\partial^\mu\phi^{(\alpha)\dagger}-m^2\phi^{(\alpha)}\phi^{(\alpha)\dagger},
\end{equation}
and $\alpha=\{0,\,1\}$. There is no constraint in this situation because $\partial_\mu\Theta=0$. Accordingly, the hamiltonian density will be
\begin{equation}\label{tc001}
\nonumber\mathcal H=\cos^2\Theta_0\,\mathcal H^{(0)}+\sin^2\Theta_0\,\mathcal H^{(1)},
\end{equation}
where
\begin{equation}
\mathcal H^{(\alpha)}=\Pi^{(\alpha)\dagger}\Pi^{(\alpha)}+\bm\nabla\phi^{(\alpha)\dagger}\bm{\cdot\nabla}\phi^{(\alpha)}+m^2\phi^{(\alpha)\dagger}\phi^{(\alpha)}.
\end{equation}
Following the quantization procedure of the $\mathbbm C$QFT, the complex components of the quaternionic fields are
\begin{equation}\label{tc002}
\widehat\phi^{(\alpha)}\big( x\big)=\int\frac{d^3k^{(\alpha)}}{\sqrt{(2\pi)^3\big|2k^{(\alpha)}_0\big|}}\Big\{\exp\big[-ik^{(\alpha)}_\mu x^\mu\big]\widehat a^{(\alpha)}+ \exp\big[ik^{(\alpha)}_\mu x^\mu\big] \widehat b^{(\alpha)\dagger}\Big\},
\end{equation}
and the commutation relations	(\ref{fc09}) and (\ref{fc11}) are satisfied in this case as well.
Further interesting results are the energy and charge operators, namely
\begin{equation}\label{tc01}
:\widehat H:=
\cos^2\Theta_0\int d^3k^{(1)}k^{(1)}_0\left[\widehat a^{(1)\dagger}\widehat a^{(1)}+\widehat b^{(1)\dagger}\widehat b^{(1)}\right]+
\sin^2\Theta_0\int d^3k^{(2)}k^{(2)}_0\left[\widehat a^{(2)\dagger}\widehat a^{(2)}+\widehat b^{(2)\dagger}\widehat b^{(2)}\right]
\end{equation}
and
\begin{equation}\label{tc02}
\widehat Q=
\cos^2\Theta_0\int d^3k^{(1)}\left[\widehat a^{(1)\dagger}\widehat a^{(1)}-\widehat b^{(1)\dagger}\widehat b^{(1)}\right]+
\sin^2\Theta_0\int d^3k^{(2)}\left[\widehat a^{(2)\dagger}\widehat a^{(2)}-\widehat b^{(2)\dagger}\widehat b^{(2)}\right].
\end{equation}
The parameter $\Theta_0$  acts as a weight between the states, a different situation from (\ref{fc17}-\ref{fc18}), were the weight of each contribution is equal because the trigonometric oscillation tends to equalize the contributions. Recalling that the complex states are such that
\begin{equation}
a^\dagger(\bm k)\big|0\big\rangle=\big|k\big\rangle\qquad\mbox{and}\qquad b^\dagger(\bm k)\big|0\big\rangle=\big|\tilde k\big\rangle.
\end{equation}
The allowed states are as follows
\begin{equation}\label{tc03}
\big|k^{(1)},\,k^{(2}\big\rangle,\qquad \big|k^{(1)},\,\tilde k^{(2}\big\rangle,\qquad\big|\tilde k^{(1)},\,k^{(2}\big\rangle,\qquad\big|\tilde k^{(1)},\,\tilde k^{(2}\big\rangle.
\end{equation}
In terms of the field operators that generate each of these states, we have
\begin{eqnarray}
\big|k^{(1)},\,k^{(2)}\big\rangle &=&\cos\Theta_0 \big|k^{(1)}\big\rangle+\sin\Theta_0\big| k^{(2)}\big\rangle \,j\,=\,\left[\cos\Theta_0 \widehat\phi^{(1)\dagger}+\sin\Theta_0\widehat\phi^{(2)\dagger}j\right]\Big| 0 \Big\rangle\\
\big|k^{(1)},\,\widetilde k^{(2)}\big\rangle &=&\cos\Theta_0 \big|k^{(1)}\big\rangle+\sin\Theta_0\big| \widetilde k^{(2)}\big\rangle \,j\,=\,\left[\,\cos\Theta_0 \widehat\phi^{(1)\dagger}+\sin\Theta_0\widehat\phi^{(2)}j\,\right]\Big| 0 \Big\rangle\\
\big|\widetilde k^{(1)},\,k^{(2)}\big\rangle &=&\cos\Theta_0 \big|\widetilde k^{(1)}\big\rangle+\sin\Theta_0\big| k^{(2)}\big\rangle \,j\,=\,\left[\,\cos\Theta_0 \widehat\phi^{(1)}+\sin\Theta_0\widehat\phi^{(2)\dagger}j\,\right]\Big| 0 \Big\rangle\\
\big|\widetilde k^{(1)},\,\widetilde k^{(2)}\big\rangle &=&\cos\Theta_0 \big|\widetilde k^{(1)}\big\rangle+\sin\Theta_0\big| \widetilde k^{(2)}\big\rangle \,j\,=\,\left[\,\cos\Theta_0 \widehat\phi^{(1)}\,+\,\sin\Theta_0\widehat\phi^{(2)}j\,\right]\Big| 0 \Big\rangle
\end{eqnarray}

The two components solution can be considered as a particular case of the previous four components case, and is a reference to compare the quaternionic and the complex solutions. The two complex components of the quaternionic field is similar to the two real components structure of the complex scalar field. We also observe that the non-associative feature of the four components quantization is also present here.
\section{\;\bf Concluding remarks\label{CR}}

In this article we presented a quantization method of quaternionic scalar field, a result that follows from the previous solution of 
the quaternionic relativistic Klein-Gordon equation \cite{Giardino:2021lov}. The decisive concept of our quantization method was to use the complex components as quantization variables, and thus avoid the difficulties in obtaining a quaternionic field theory using a pure quaternionic formalism, where the definition of a quaternionic derivative would be unavoidable.

As predicted, the quaternionic fields encompass a higher number of degrees of freedom in comparison to the complex case, and thus establish a
more flexible field theory in order to describe potentially more sophisticated physical phenomena. Additionally, the quaternionic theory presents a non-associative feature, and hence physical theories where such kind of property is required can be described by a quaternionic theory. The physical processes that would be described using such a quaternionic formalism is a matter of debate, mainly because there are still open theoretical questions that must be solved. First of all, it is necessary to understand a way to introduce interaction terms in the theory, and also how to interpret this new theory. Another directions of future research are the development of a quaternionic Dirac field theory,  the quantization of the quaternionic electromagnetic field \cite{Giardino:2020uab}, and the analysis of the Lorentz invariance \cite{Rigolin:2020vaq}, or even the expression of this theory in terms of complex quaternions \cite{DeLeo:1995ww}.
Our results indicate that quaternions can be quintessential in order to build wave functions composed by two components of different energies. If such system will play a significant role in the Physics of the future, we will finally know the kind of problem quaternions were suitable to solve.

\paragraph{Data availability statement}The author declares that data sharing is not applicable to this article as no datasets were generated or analysed during the current study.

%
%
%
%
\begin{footnotesize}

\end{footnotesize}

\end{document}